\documentclass{PoS}

\usepackage{cite}
\usepackage{amsmath}
\usepackage{amssymb}
\usepackage{mathrsfs}
\usepackage{graphics}
\usepackage{subfig}
\usepackage{xspace}
\newcommand{\vecb}[1]{\textit{\textbf{#1}}}
\newcommand{\eq}[1]{eq.~\eqref{eq:#1}}
\newcommand{\Eq}[1]{Eq.~\eqref{eq:#1}}

\newcommand{\fig}[1]{fig.~\ref{fig:#1}}

\newcommand{\df}{\mathrm{d}}
\newcommand{\Eventtwo}{\texttt{Event2}\xspace}
\newcommand{\Artemide}{\texttt{arTeMiDe}\xspace}

\title{
\vspace*{-1.2cm}
\begin{minipage}{\textwidth}
{\normalfont\small Nikhef 2019-028}\\
\end{minipage}\\[10pt]
Studying transverse momentum distributions with jets at N$^3$LL}

\ShortTitle{Studying TMDs with jets at N$^3$LL}
\author{Daniel Gutierrez-Reyes,$^{a}$ Ignazio Scimemi,$^{a}$ Wouter J.~Waalewijn$^{bc}$ and \speaker{Lorenzo Zoppi}$^{bc}$\\
        \llap{$^a$}Departamento de F\'isica Te\'orica, Universidad Complutense de Madrid (UCM) and IPARCOS,\\
        E-28040 Madrid, Spain \\
        \llap{$^b$}Institute for Theoretical Physics Amsterdam and Delta Institute for Theoretical Physics, University of Amsterdam, \\
        Science Park 904, 1098 XH Amsterdam, The Netherlands\\
        \llap{$^c$}Nikhef, Theory Group,\\
        Science Park 105, 1098 XG, Amsterdam, The Netherlands\\
        E-mail: \email{dangut01@ucm.es}, \email{ignazios@fis.ucm.es}, \email{w.j.waalewijn@uva.nl}, \email{l.zoppi@uva.nl}}
\abstract{Semi-inclusive deep inelastic scattering (SIDIS) is a promising channel for the extraction of transverse momentum dependent distributions at future colliders. In this context, we recently developed a framework that uses jets (instead of single hadrons) to achieve reduced sensitivity to final-state non-perturbative effects. A suitable non-standard jet definition allows us to apply the factorization formulas valid for hadrons to jets of arbitrary size, by just replacing fragmentation functions with the jet functions we computed. Besides presenting the framework, we will show numerical predictions at N$^3$LL accuracy.}
\FullConference{XXVII International Workshop on Deep-Inelastic Scattering and Related Subjects - DIS2019\\
		8-12 April, 2019\\
		Torino, Italy}
		
\begin{document}
\section{Motivation}
The structure of the proton has been a fundamental research topic for a long time. Besides intrinsic interest in the complex interplay of gluons and quarks in a bound state, an increasingly detailed description of the nucleon is demanded to achieve high precision at hadron colliders.
From the early days of the parton model, the main focus has been on distributions in longitudinal momentum fraction of quarks and gluons extracted from hadrons (Parton Distribution Functions, PDFs), or fragmenting into hadrons (Fragmentation Functions, FFs). Transverse Momentum dependent Distributions (TMDs) generalize this picture to the transverse plane, allowing to describe processes where the transverse momentum of final-state objects is measured. Relevant examples are the di-hadron relative transverse momentum in $e^+ e^-$ collisions, the transverse momentum of a $\gamma^*/Z$ boson in $pp$ collisions, and the hadron transverse momentum in SIDIS.

The key ingredient to study nucleons via high-energy experiments is factorization: low-energy, non-perturbative physics is confined in hadronic matrix elements, that are universal and fitted to data, while the high-energy, process-dependent physics is still described by perturbative quantum field theory. 
In the case of SIDIS, the cross section differential in transverse momentum schematically factorizes as the product of a hard function, a TMD PDF and a TMD FF.
Although SIDIS is promising for the extraction of TMDs at electron-proton colliders, it gets complicated by the presence of non-perturbative physics in both the initial and final state of the process. Measuring a perturbatively calculable jet instead of a hadron would in principle remove the main source of final-state non-perturbative uncertainty, allowing for a cleaner extraction of the TMD PDFs. This would be particularly interesting at the future Electron-Ion Collider (EIC). 

With this motivation in mind, we investigated what happens to TMD factorization when jets are measured instead of hadrons~\cite{Gutierrez-Reyes:2018qez}. There we proved that a class of processes obey the same factorization formulas valid for single hadrons, with the TMD FF replaced by a TMD jet function we computed at NLO. 
With a recoil-free jet axis such as the Winner-Take-All (WTA,~\cite{Bertolini:2013iqa}), this works independent of the jet radius, providing a simple and comprehensive framework.
In ref.~\cite{Gutierrez-Reyes:2019vbx} we supplemented our analysis with numerical predictions for $e^+e^- \rightarrow \mbox{dijet}$ and SIDIS. We found that even for moderate values of the jet radius parameter $(R\gtrsim 0.5)$ using the large-$R$ limit of the jet function gives an excellent approximation of the TMD cross section. We then determined numerically the NNLO jet function in this limit, achieving N$^3$LL accuracy. In the following, we will summarize our framework and numerical predictions, and comment on the status of our analysis.

\section{Framework}	\label{sec:framework}
For definiteness we focus on DIS with a jet in the final state, but our formalism easily extends to e.g. $e^+e^-$ collisions. We work in the Breit frame and define the transverse momentum as
\begin{align}
  \vecb{q} = \vecb{P}_J /z + \vecb{q}_{\rm in}\, ,
\end{align}
i.e. as the jet transverse momentum with respect to the beam rescaled by the jet energy fraction $z$, relative to the transverse momentum $\vecb{q}_{\rm in}$ of the incoming quark in the proton. Other relevant variables are the Bjorken $x$ and the virtuality of the electroweak boson, $Q^2 > 0$. The latter sets the scale of the hard partonic interaction, so factorization requires $q_T = |\vecb{q}| \ll Q$. In fact, the angle
\begin{align}
  \theta = \arctan\big(2q_T/Q\big) \simeq 2q_T/Q \ll 1
\end{align}
is the natural power counting parameter for the factorization analysis. This small quantity can compete with the jet size $R$, and we consider in turn the hierarchies $R \sim \theta$ and $R \gg \theta$. The case $R \ll \theta$ provides an insightful consistency check with fragmentation, but is of limited practical interest, and we point the interested reader to ref.~\cite{Gutierrez-Reyes:2019vbx} for its study.

In Soft-Collinear Effective Theory, factorization occurs in the Lagrangian from separation of modes, whose scaling is set by kinematics~\cite{Bauer:2001yt}. A generic four-vector $p$ decomposes along light-cone reference vectors $n$, $\bar{n}$ as $p^-\! = \bar{n}\cdot p,\,p^+\! = n \cdot p,\,\vecb{p}$. If $\theta \sim R$, the relevant modes scale as
\begin{align} \label{eq:scaling}
  \mbox{collinear: } p_c \sim (1,\theta^2,\theta)Q, \qquad \quad \mbox{soft: }p_s \sim (\theta, \theta, \theta)Q
\end{align}
and the factorization formula up to $\mathcal{O}(q_T^2/Q^2)$ power corrections reads
\begin{align} \label{eq:factorization}
  \frac{{\rm d} \sigma_{ep \rightarrow e\, \rm{jet}}}{{\rm d} Q^2\, {\rm d} x \, {\rm d} z\, {\rm d}\vecb{q}} = \sum_{f=q,\bar{q}}
  H_{ef\rightarrow ef} (Q^2,x) \int \frac{{\rm d}^2 \vecb{b}}{(2\pi)^2} e^{-{\rm i}\, \vecb{b}\,\cdot \vecb{q}} F_{p\rightarrow f} (x,\vecb{b},\mu,\zeta)\, J^{\rm axis}_f (z,\vecb{b},\tfrac{QR}{2},\mu,\zeta)\, .
\end{align}
Here $H$ is the hard function, including tree-level DIS cross section, $F$ is the TMD PDF, and $J$ is the jet function, which in general depends on the choice of axis, standard or WTA. The formula has a simpler form in terms of the impact parameter $\vecb{b}$, Fourier conjugate of $\vecb{q}$. As is common in TMD calculations, the functions are affected by rapidity divergences that are not fixed by dimensional regularization, but require an additional regulator and ultimately cancel between soft and collinear function~\cite{Collins:1984kg,Becher:2010tm,Chiu:2011qc,Ji:2004wu,Scimemi:2017etj}. In fact,
in \eq{factorization} we reabsorbed the soft function in the TMD PDF and jet function~\cite{GarciaEchevarria:2011rb,Echevarria:2015uaa}. The procedure introduces a rapidity scale $\zeta$ in addition to the UV scale $\mu$.

The reason why \eq{factorization} holds for jets is that the scaling in \eq{scaling} is the same as for hadrons. Intuitively, if $R \sim \theta \ll 1$ the jet is too narrow to contain a sensible amount of soft radiation, thus it still recoils against the overall soft radiation, as a single hadron would.
We expect the situation to change in the regime $R \gg \theta$, where the jet does contain a large amount of soft radiation. This happens dramatically with standard jets: by construction the axis balances the transverse momentum of all the constituents of the jet, so the latter recoils only against \emph{external} soft radiation. Emissions inside the jet are hard~\cite{Becher:2016mmh,Larkoski:2015zka} and induce soft Wilson lines, making the observable highly sensitive to non-global logarithms~\cite{Dasgupta:2001sh}. Surprisingly, instead, with the WTA axis nothing changes. By construction the axis tracks the collinear radiation, and again the jet recoils against the \emph{overall} soft radiation. \Eq{factorization} is still valid, and in addition the jet function largely simplifies,
\begin{align} \label{eq:largeR}
  J^{\rm WTA}_f (z,\vecb{b},\tfrac{QR}{2},\mu,\zeta)  = \delta(1-z) \mathscr{J}_f^{\rm WTA}(\vecb{b},\mu,\zeta)\,
  \big[ 1 + \mathcal{O}\big(\theta^2/R^2\big)\big]\, .
\end{align}
All the radiation is now enclosed in the jet, so the dependence on the jet size and energy fraction becomes trivial. Since the remaining arguments are constrained by evolution, $\mathscr{J}$ is known up to a constant. We exploited this to extract numerically the two-loop constant using \Eventtwo.

As a consequence of \eq{factorization}, consistency under Renormalization Group flow imposes that our jet functions have the same double-scale evolution as the TMDs,
\begin{align} \label{eq:RGE}
  (\df/\df \mu) J_f (\vecb{b},\mu,\zeta) = \gamma_f (\mu,\zeta)\, J_f (\vecb{b},\mu,\zeta)\, , \quad
  (\df/\df \zeta) J_f (\vecb{b},\mu,\zeta) = - \mathcal{D}_f(\vecb{b},\mu)\, J_f (\vecb{b},\mu,\zeta)\, ,
\end{align}
where we omitted non-relevant arguments. This is convenient, since the rapidity anomalous dimension $\gamma_f$ and UV anomalous dimension $\mathcal{D}_f$ are known up to three loops~\cite{Moch:2005tm,Baikov:2009bg,Li:2016ctv},
providing all the necessary ingredients for N$^3$LL resummation of large logarithms $\ln(q_T/Q)$. The recently proposed $\zeta$-prescription~\cite{Scimemi:2018xaf} sensibly sets the initial rapidity scale $\zeta_0$ as a function of $\mu_0$ so as to cancel any dependence of the (jet) TMDs on $\mu_0,\zeta_0$. 
Defining TMDs in this way disentangles the uncertainty related to evolution from the non-perturbative uncertainty, and is therefore theoretically preferred.

\section{Numerical predictions}	
\begin{figure*}[tb]
  \centering
  \includegraphics[width=0.48\textwidth]{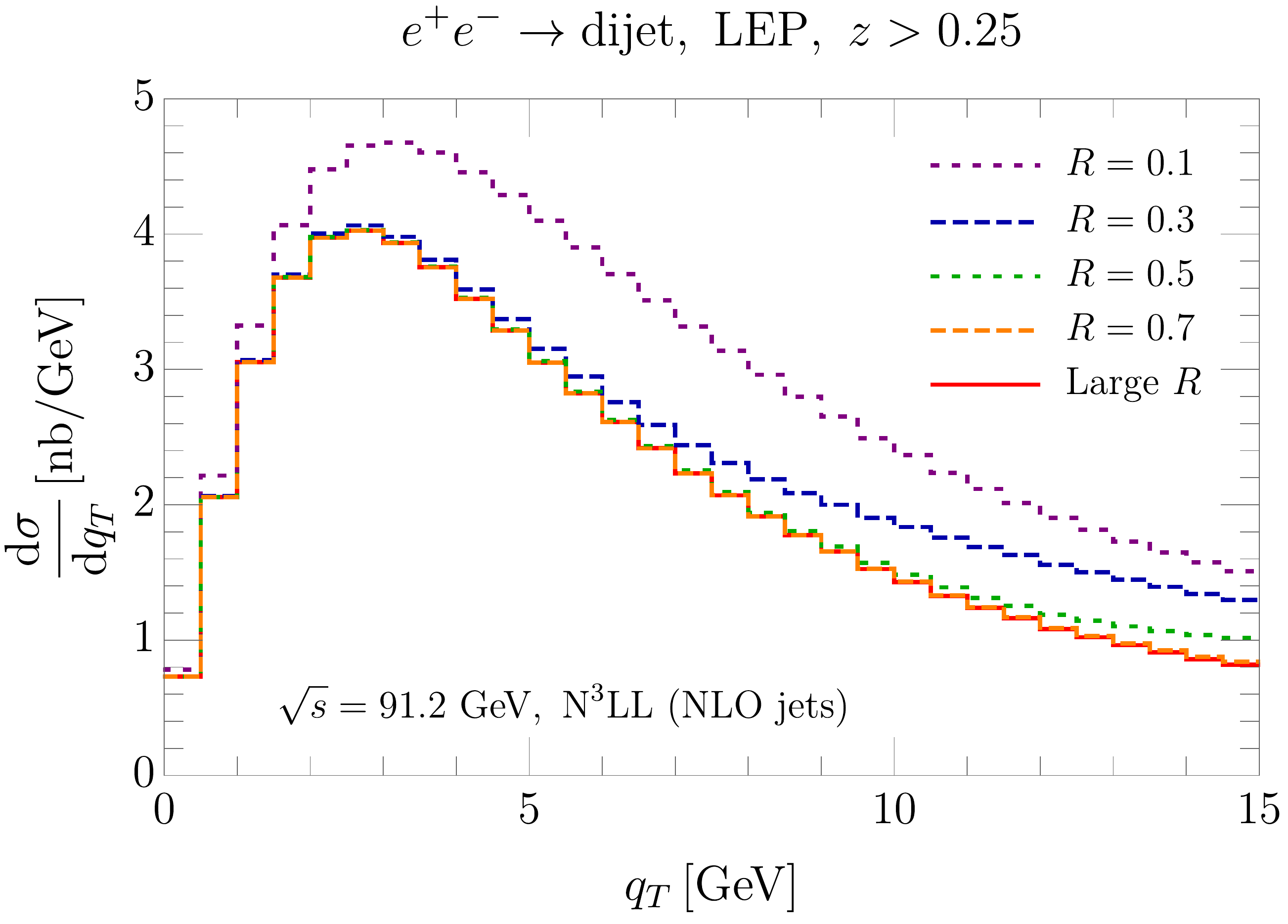} \quad
   \includegraphics[width=0.48\textwidth]{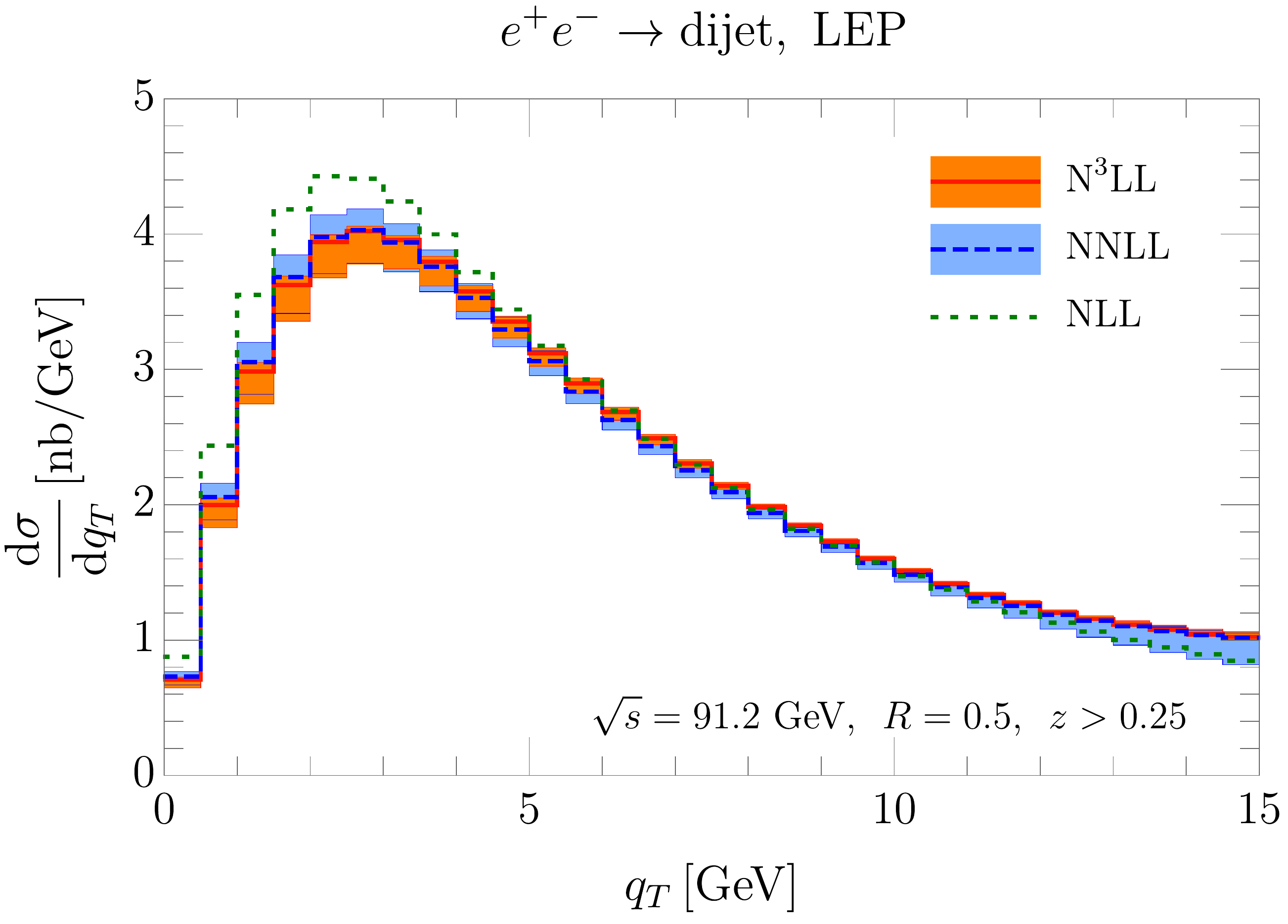} 
  \caption{Differential transverse momentum distribution for dijet production at LEP. (\emph{Left}) predictions for different radii and the large $R$ limit for comparison. (\emph{Right}) convergence of the perturbative series for $R = 0.5$. The NLL band within the $\zeta$-prescription is artificially small and not shown.}
\label{fig:e+e-}
\end{figure*}

Based on the framework above, we modified the code \Artemide\footnote{Web-page: https://teorica.fis.ucm.es/artemide/
  		Repository: https://github.com/VladimirovAlexey/artemide-public} to get predictions with jets. First we considered $e^+e^- \rightarrow \mbox{dijet}$, 
with \fig{e+e-} showing predictions for LEP. We plot the cross section differential in $q_T$ in the region where power corrections are suppressed, imposing a cut $z > 0.25$ over the jet energy fractions. The left panel shows predictions for different jet radii, and the large $R$ limit from \eq{largeR} for comparison. The $R \gg \theta$ regime is an excellent approximation, with $R = 0.7$ almost coinciding with the large $R$ limit and $R = 0.5$ showing small deviations only in the tail. This motivates us to include corrections from the large-$R$, two loop expression we numerically extracted for the function $\mathscr{J}^{\rm WTA}$.
The right panel of \fig{e+e-} shows the convergence of the perturbative series, where we obtain theoretical error bands by varying of a factor 2 (0.5) the factorization scales around the central values $\mu = \sqrt{s}$, $\mu_0 = 2e^{-\gamma_E}/b_T + 2 \mbox{GeV}$. As a check of the framework, the leap from NLL to NNLL reduces from NNLL to N$^3$LL and the error band shrinks.

\begin{figure*}[tb]
  \centering
  \includegraphics[width=0.48\textwidth]{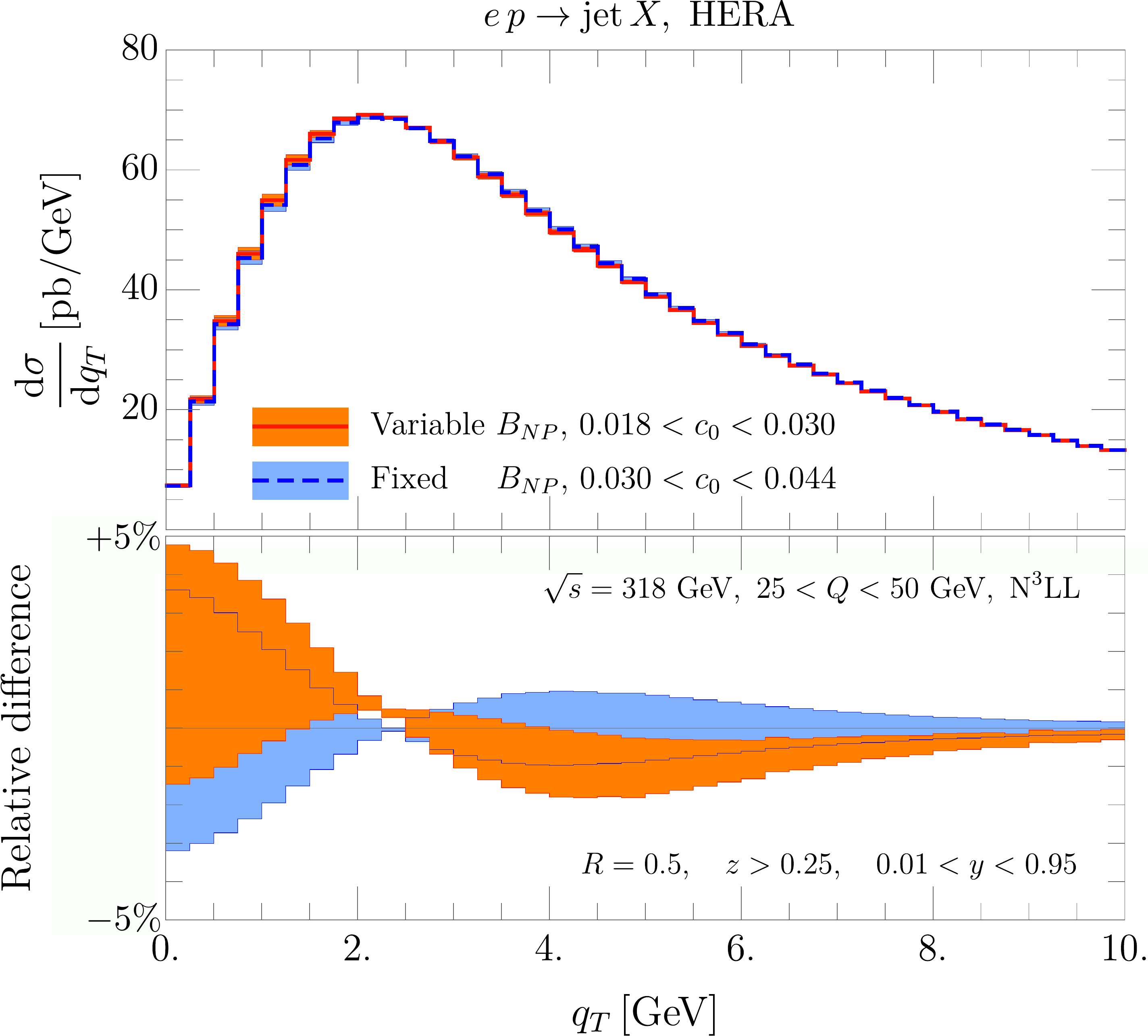} \quad
   \includegraphics[width=0.48\textwidth]{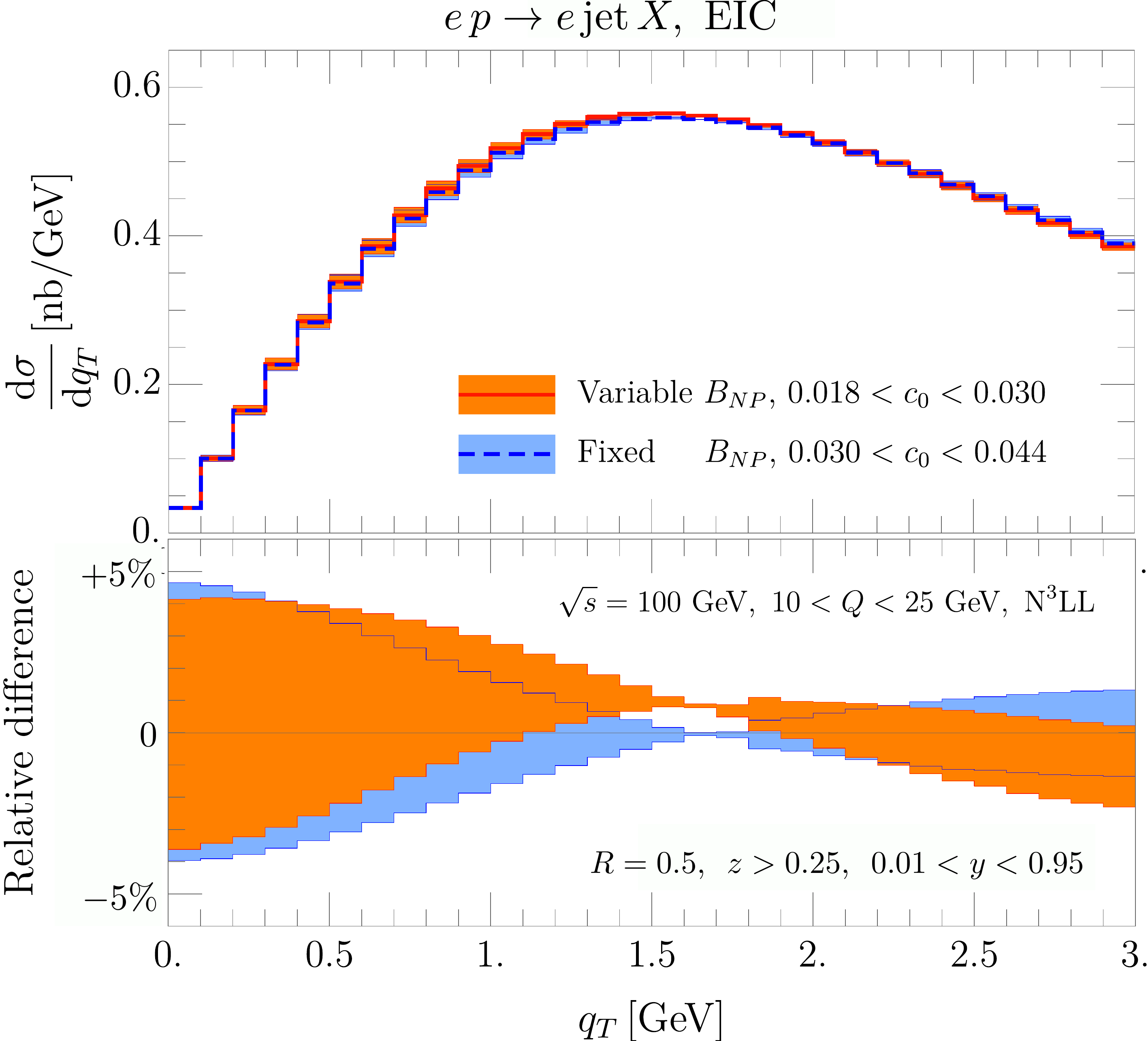} 
  \caption{Differential transverse momentum distribution for SIDIS with jets at HERA (left) and EIC (right), for which we set $\sqrt{s} = 100$ GeV. We estimate the sensitivity to non-perturbative initial state physics from the variation of the dominant parameter $c_0$ in the model we take from~\cite{Bertone:2019nxa}.}
\label{fig:SIDIS}
\end{figure*}

Finally, we obtained predictions for SIDIS experiments, with a jet in the final state. We considered the center of mass energies of HERA ($\sqrt{s} = 318$ GeV) and an hypothetic $100$-GeV EIC. Beside predicting the shape of the transverse momentum distribution, we are interested in estimating the sensitivity to initial-state non-perturbative physics. We did so within the non-perturbative model of ref.~\cite{Bertone:2019nxa} for TMD PDFs, varying the parameters within their fit uncertainties. One of these, $c_0$, dominates the uncertainty, so its variation alone gives a reliable description of the magnitude of the effect. In \fig{SIDIS} we show the results integrating over jet energy fraction $(z > 0.25)$, elasticity $(0.01 < y < 0.95)$ and virtuality ($25 < Q < 50$ GeV for HERA, $10 < Q < 25$ GeV for EIC) of the process. We find for both configurations that the size of the variation is mild ($\sim 5 \%$ at low $q_T$, a few percent elsewhere) but similar to using hadrons in the final state of the process.

\section{Outlook}	
Our analysis shows that recoil-free jets are a valid alternative to single hadrons when final-state transverse momenta are measured. They benefit from a robust factorization and evolution framework whose ingredients are known up to N$^3$LL and they have reduced sensitivity to final-state non-perturbative physics. The main question still to be answered is whether this smaller non-perturbative uncertainty compensates for new sources of error introduced by the jet measurement. For instance, the position of the jet axis is less precisely measured than a single charged hadron, where the angular resolution of the tracker is superior. Defining the jets on only charged hadrons would limit this issue, but this could in principle increase the sensitivity to non-perturbative physics. In addition, the results depend on the considered range of energy fraction, $z>z_{\rm cut}$, which also enter the definition of transverse momentum $\vecb{q} = \vecb{P}_J/z$. This should be marginal since we showed the large-$R$ limit, where all the energy goes into the jet, to be a solid approximation for the finite-$R$ case. The size of these effects could be assessed with a dedicated Monte Carlo study. Preliminary results look promising, suggesting that the additional sources of error are under control.
Finally, we mention that an alternative possibility to limit final-state hadronization effects would be using groomed jets. This is the subject of a forthcoming work~\cite{Gutierrez-Reyes:2019msa}.

\bibliographystyle{JHEP}
\bibliography{refs.bib}
\end{document}